\pdfoutput=1
\documentclass[11pt]{article}

\usepackage[final]{acl}

\usepackage{times}
\usepackage{latexsym}
\usepackage{microtype}
\usepackage{graphicx}

\usepackage{booktabs} 
\usepackage{multirow,tabularx,multicol}
\usepackage[table]{xcolor}
\definecolor{verylightgray}{gray}{0.9}
\definecolor{verylightred}{rgb}{1,0.9,0.9}
\definecolor{aliceblue}{rgb}{0.94,0.97,1.0}
\usepackage{hyperref}
\usepackage{subcaption}

\usepackage{amsmath}
\usepackage{amssymb}
\usepackage{mathtools}
\usepackage{amsthm}
\usepackage{amsfonts}

\usepackage{paralist}

\usepackage[T1]{fontenc}

\usepackage[utf8]{inputenc}

\usepackage{inconsolata}

\usepackage{algorithm}
\usepackage{pifont}

\usepackage{algorithmicx} %
\usepackage{algpseudocode} %

\usepackage{arydshln}
\newcommand{\bc}[2][1.0]{%
  \scalebox{#1}{\ding{\numexpr181+#2\relax}}%
}

\usepackage{newfloat}
\usepackage{listings}
\DeclareCaptionStyle{ruled}{labelfont=normalfont,labelsep=colon,strut=off} 
\floatstyle{ruled}
\newfloat{listing}{tb}{lst}{}
\floatname{listing}{Listing}
\title{Rhythm-TTS: High-Fidelity Shortcut Latent Modeling and Probabilistic Rhythm Generation for Zero-Shot Text-to-Speech}

\author{
 \textbf{Hieu-Nghia Huynh-Nguyen\textsuperscript{1, 4}} \quad
 \textbf{Ngoc-Son Nguyen\textsuperscript{1}} \quad
 \textbf{Long-Khanh Pham\textsuperscript{2}} \quad \\
 \textbf{Huynh Nguyen Dang\textsuperscript{1}} \quad 
 \textbf{Huy Tien Nguyen\textsuperscript{3, 4}} \quad
 \textbf{Van Nguyen\textsuperscript{1}\thanks{Corresponding author}}
\\
 \textsuperscript{1}Quantum AI \& Cyber Security Institute, FPT Corporation, Vietnam \\
 \textsuperscript{2}FPT Software AI Center, FPT Corporation, Vietnam \\
 \textsuperscript{3}Faculty of Information Technology, University of Science, Ho Chi Minh City, Vietnam \\
 \textsuperscript{4}MoMo, Vietnam \\
 \texttt{huynhnguyenhieunghia1999@gmail.com, vannth19@fpt.com}
}

\begin{document}
\maketitle
\begin{abstract}
Zero-shot text-to-speech (TTS) systems are commonly built on either discrete codec tokens or continuous acoustic representations. Discrete-token models benefit from stable sequence modeling and explicit speech factorization, but they ultimately synthesize speech through codebook lookup and codeword composition, which can accumulate errors and impose a quality ceiling on naturalness and speaker similarity. Continuous-representation models can generate fine-grained speech latents directly, but often require heavier in-context modeling and less explicit alignment. This paper presents \textbf{Rhythm-TTS}, a shortcut zero-shot TTS framework that combines the factorizability of discrete codec tokens with the fidelity of continuous latent generation. Instead of using predicted codec tokens as the final acoustic representation, Rhythm-TTS uses them as a semantic-acoustic prior and directly generates holistic continuous speech latents with flow matching, bypassing the discrete-to-latent reconstruction path at inference time. We further address an underexplored dimension of speech naturalness: rhythm diversity across multiple synthesis runs. Motivated by the hypothesis that rhythm variation is largely governed by phoneme duration changes and the occurrence of silent pauses, we introduce probabilistic Duration and Silence Generators to sample dynamic pacing patterns before acoustic generation. Experiments on zero-shot TTS benchmarks evaluate intelligibility, naturalness, speaker similarity, efficiency, and run-to-run rhythm diversity, showing that the proposed shortcut latent modeling and probabilistic rhythm generation improve the fidelity and temporal naturalness of non-autoregressive zero-shot TTS. Code and audio samples are available at our demo page\footnote{\url{https://rhythm-tts.github.io}}.
\end{abstract}

\section{Introduction}
Given a target text transcript and a speech prompt from an unseen speaker, zero-shot TTS generates a speech waveform that preserves the prompt speaker identity while speaking the target text. Recent systems have made strong progress in intelligibility, naturalness, and speaker similarity, largely due to neural audio codecs, large-scale sequence models, diffusion models, and flow matching. However, the dominant modeling choices still expose a trade-off between stability and fidelity. Discrete codec tokens provide a compact and stable target for language-model-style generation, while continuous acoustic representations preserve fine-grained information that is important for naturalness and speaker identity.

A large family of zero-shot TTS systems models speech as a sequence of discrete tokens obtained from neural codecs \cite{soundstream, encodec}. Systems such as VALL-E and its variants \cite{valle, valle2, valle-x, valle-r, ella-v, voicecraft, sparktts} benefit from the robustness of discrete sequence modeling, and recent diffusion- or flow-based systems also use codec tokens to simplify the acoustic generation problem \cite{ns3, oz}. Nevertheless, a discrete-token pipeline has an inherent bottleneck: the target continuous speech latent is approximated by selecting and combining representative codewords from finite codebooks. Even when the token sequence is correct, codebook lookup can discard fine acoustic details, while token prediction errors can be propagated through the reconstruction path. This makes purely discrete-token generation potentially suboptimal for the aspects most sensitive to fine-grained acoustics, especially naturalness and speaker similarity.

Continuous-representation models \cite{voicebox, e2tts, f5tts} avoid this codebook bottleneck by generating mel-spectrograms or latent acoustic representations directly. They often achieve high perceptual quality because the model is not forced to express speech only through finite codec indices. However, these models usually rely on heavy in-context modeling to learn text-speech alignment and prompt conditioning implicitly, which can demand large training data and expensive backbones. This motivates a hybrid question: can we keep the stability and explicit structure of discrete codec tokens while avoiding the final discrete-to-continuous reconstruction bottleneck?

We answer this question with \textit{Rhythm-TTS}, a shortcut latent modeling framework for zero-shot TTS. The core idea is to treat generated discrete codec tokens as an intermediate prior rather than the final acoustic representation. A compact Code Generator first predicts multi-level codec tokens conditioned on the input phonemes and the speech prompt. Instead of decoding these tokens through the codec codebooks as in a cascade pipeline, a flow-matching Denoiser uses their embeddings as a semantic-acoustic starting point and directly generates the holistic continuous speech latents consumed by the codec decoder. This shortcut formulation preserves the stable planning role of discrete tokens while eliminating the codeword-lookup reconstruction path during synthesis.

Beyond fidelity, Rhythm-TTS focus on a second underexplored aspect of zero-shot TTS: \textit{rhythm diversity}. Human speakers naturally vary phoneme durations and insert pauses differently across repeated readings of the same sentence. Autoregressive TTS models may exhibit such variation due to stochastic token generation, but practical non-autoregressive systems often depend on deterministic duration predictors and length regulators \cite{fastspeech}. As a result, repeated synthesis can produce nearly fixed timing patterns even when acoustic details vary. We hypothesize that run-to-run rhythm diversity is mainly driven by two factors: changes in phoneme durations and the appearance, placement, and length of silent segments. Based on this hypothesis, we introduce probabilistic Duration and Silence Generators, which sample both spoken-phoneme durations and optional pause durations before acoustic generation.

This paper makes the following contributions:
\begin{itemize}
    \item We propose \textbf{shortcut latent modeling} for zero-shot TTS: discrete codec tokens are generated for stable semantic-acoustic planning, but the final speech is synthesized from continuous holistic latents generated end-to-end by flow matching.
    \item We introduce a \textbf{discrete-conditioned continuous Denoiser} that starts from a codec-token-derived prior and directly regresses toward target continuous speech latents, avoiding the cascade of token prediction followed by codeword reconstruction.
    \item We propose \textbf{probabilistic rhythm generation} by jointly modeling phoneme durations and silent pauses through complementary Duration and Silence Generators. This enables non-autoregressive zero-shot TTS to synthesize diverse speech rhythms by varying both phoneme timing and pause placement across repeated generations, increasing run-to-run variation in phoneme timing and pause structure.
    \item We perform a comprehensive evaluation of intelligibility,
    naturalness, speaker similarity, efficiency, and rhythm diversity. Controlled experiments show that shortcut generation substantially improves naturalness and speaker similarity while preserving WER. Explicit duration and silence sampling increases diversity by $2.69\times$ over fixed-rhythm generation without increasing WER and achieves the best WER-normalized diversity metric.
\end{itemize}
\section{Related Work}
\textbf{Discrete-token zero-shot TTS.} Neural codecs convert speech waveforms into compact token sequences through vector quantization \cite{soundstream, encodec}. This representation has enabled language-model-style zero-shot TTS, where speech generation is formulated as predicting codec tokens conditioned on text and a speech prompt. VALL-E \cite{valle} and its successors \cite{valle2, valle-x, valle-r, ella-v, voicecraft, sparktts} demonstrate that discrete acoustic tokens can provide strong stability and scalability. Diffusion-based token models such as NaturalSpeech 3 \cite{ns3} and related prior-generation methods \cite{oz} further show that codec token hierarchies can expose useful speech factors. Our work differs by not treating generated discrete tokens as the final acoustic representation. Instead, we use them as a prior for continuous latent generation, eliminating reliance on the codec codebook reconstruction path.

\textbf{Continuous acoustic generation.} Continuous-representation TTS models, including diffusion- and flow-matching-based systems \cite{kang23_interspeech, tran23d_interspeech, ns2, kim2023pflow, matcha-tts, e2tts, f5tts}, generate mel-spectrograms or acoustic latents directly. These systems can preserve fine-grained acoustic details and often achieve strong naturalness. However, direct continuous generation usually requires the model to learn text alignment, prompt conditioning, and acoustic realization jointly. Rhythm-TTS keeps the continuous generation target but introduces a discrete-token-derived shortcut prior, allowing the continuous generator to focus on refining holistic latents rather than learning the entire generation process from noise or masked acoustic context.

\textbf{Efficient and shortcut generative modeling.} Efficiency has been studied from several angles. Small-E \cite{small-e} replaces Transformer blocks with recurrent or linear-attention alternatives such as RWKV \cite{rwkv}, Mamba \cite{mamba}, and Gated Linear Attention \cite{gatedlinearattentiontransformers}. MobileSpeech \cite{mobilespeech} targets compact and parallel generation for mobile zero-shot TTS. Other works improve the sampling process: ZipVoice \cite{zipvoice} uses Zipformer \cite{zipformer} and flow distillation; ShallowFlow \cite{shallowflow} initializes generation from intermediate states; RapFlow-TTS \cite{rapflow} adopts consistency flow matching. One-step, consistency, and shortcut models \cite{cm-tts, comospeech} reduce the number of generative transitions in broader speech generation settings. Rhythm-TTS uses the term \textit{shortcut} in a complementary sense: it shortcuts the discrete-token cascade by directly generating continuous latents from a token-derived prior.

\textbf{Rhythm diversity.} Many non-autoregressive TTS systems inherit deterministic duration prediction from FastSpeech-style length regulation \cite{fastspeech}. This design is efficient and improves alignment, but it tends to produce fixed timing patterns for the same text. Probabilistic duration modeling has been explored in systems such as VITS-style models \cite{vits, vits2} and dedicated duration models \cite{probdur}. However, phoneme duration alone does not fully explain natural rhythm because silence placement and pause duration also strongly affect perceived pacing. Our probabilistic rhythm generation module augments conventional phoneme Duration Generation with a novel Silence Generator. Together, these modules enable stochastic varying of both phoneme durations and silent pause placement across runs,  thereby more faithfully reproducing the natural rhythm variations observed in repeated human speech.

\section{Method}
\label{sec:method}
\begin{figure*}[htbp]
    \centering
    \includegraphics[width=1\textwidth]{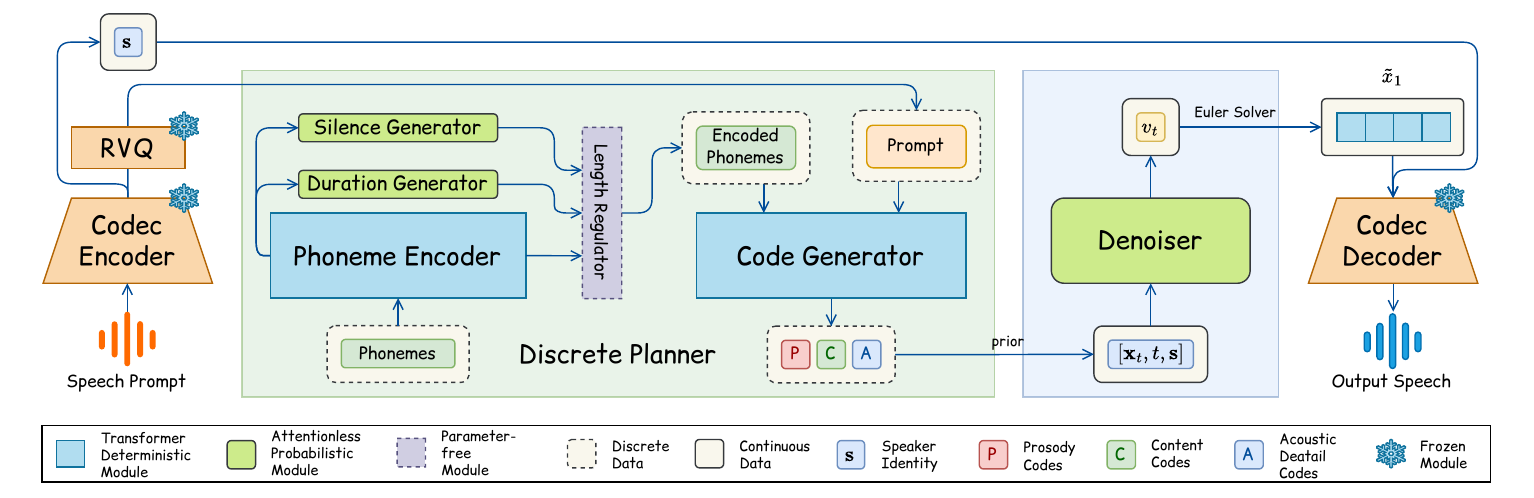}
    \caption{Overview of Rhythm-TTS. A speech prompt is encoded into speaker identity, continuous codec latents, and multi-level discrete codec tokens. The input text is first converted into a phoneme sequence, then expanded according to sampled phoneme durations and augmented with sampled silent pauses by the Probabilistic Rhythm Planner. A Discrete Planner predicts discrete codec tokens as a stable semantic-acoustic plan. Instead of reconstructing speech only by codebook lookup, Rhythm-TTS maps the generated tokens into a continuous prior and uses a flow-matching Denoiser to directly generate holistic speech latents for the codec decoder.}
    \label{fig:overall-architecture}
\end{figure*}

Rhythm-TTS is built around two design principles. First, discrete codec tokens should be used for what they are good at: stable, structured speech planning. Second, the final acoustic representation should remain continuous, because naturalness and speaker similarity depend on fine-grained details that are difficult to express with finite codebooks. Fig.~\ref{fig:overall-architecture} illustrates the resulting top-down pipeline.

\subsection{Problem Setup and Representations}
\label{sec:representations}
Let $\mathcal{P}$ be the input phoneme sequence. We use a neural codec to extract two complementary representations from speech: discrete multi-level codec tokens and continuous holistic speech latents. The prompt provides codec token sequences $\mathbf{P}^{1:6}$ and speaker identity $s$, while the target training utterance provides the continuous latent target $\mathbf{x}_1$.

Most discrete-token systems first predict target codec tokens and then reconstruct speech by looking up and combining codebook entries. We refer to this as a \textit{cascade} path. Rhythm-TTS instead uses the predicted tokens only to construct a prior $\mathbf{x}_{\mathrm{pr}}$, then directly generates the target continuous latent $\tilde{\mathbf{x}}_1$. The overall pipeline can be mathematically expressed as:
\begin{align}
\label{eq:framework}
    \mathbf{Q}^{1:6} &= \mathcal{G}_{\psi}(\mathcal{P}, \mathbf{P}^{1:6}), \\
    \mathbf{x}_{\mathrm{pr}} &= \mathcal{E}_{\mathrm{fold}}(\mathbf{Q}^{1:6}), \\
    \tilde{\mathbf{x}}_1 &= \mathcal{D}_{\theta}(\mathbf{x}_{\mathrm{pr}}, s),
\end{align}
where $\mathcal{G}_{\psi}$ is the Code Generator, $\mathcal{E}_{\mathrm{fold}}$ maps discrete-token embeddings into a continuous prior, and $\mathcal{D}_{\theta}$ is the flow-matching Denoiser.

\subsection{Probabilistic Rhythm Planner}
\label{sec:prob-dur-sil}
Before generating codec tokens, Rhythm-TTS samples a timing plan for the input phoneme sequence. Let $\mathbf{h}_{1:L} \in \mathbb{R}^{L \times D}$ be the encoded phoneme sequence. A deterministic duration predictor would output a fixed duration for each phoneme, which limits run-to-run rhythm diversity. We instead model spoken-phoneme duration as a conditional flow. Given a noise $\mathbf{d}_0 \sim \mathcal{N}(0, I)$ and target log durations $\mathbf{d}_1$, the Duration Generator is trained with:
\begin{equation}
    \label{eq:durgen}
    \mathcal{L}_{dur}(\phi) = \mathbb{E}_{\mathbf{d}_0,\mathbf{d}_1,t}
    \left\lVert
        \mathcal{D}_\phi(\mathbf{d}_t,\mathbf{h}_{1:L},t) - (\mathbf{d}_1 - \mathbf{d}_0)
    \right\rVert^2,
\end{equation}
where $\mathbf{d}_t = t\mathbf{d}_1 + (1-t)\mathbf{d}_0$ and $t \sim \mathcal{U}(0,1)$.

Silence is modeled as a separate probabilistic process because pauses are not merely long phonemes: they can appear or disappear depending on phrasing. The Silence Generator predicts optional pause durations after phonemes:
\begin{equation}
    \label{eq:silgen}
    \mathcal{L}_{sil}(\xi) = \mathbb{E}_{\mathbf{s}_0,\mathbf{s}_1,t}
    \left\lVert
        \mathcal{S}_\xi(\mathbf{s}_t,\mathbf{h}_{1:L},t) - (\mathbf{s}_1 - \mathbf{s}_0)
    \right\rVert^2,
\end{equation}
where $\mathbf{s}_t = t\mathbf{s}_1 + (1-t)\mathbf{s}_0$ and $\mathbf{s}_0 \sim \mathcal{N}(0, I)$. During sampling, spoken phonemes are constrained to have positive duration, while inserted silences are allowed to have zero duration. This design lets the model complementarily vary both speaking rate and pause structure while preserving the text order.

\subsection{Discrete Code Generator}
\label{sec:code-generator}
After duration expansion and optional silence insertion, the expanded phoneme representation is passed to a hierarchical Code Generator. The Code Generator predicts six codec-token levels, corresponding to prosody, content, and acoustic-detail code streams. At each level, prompt tokens are injected so that the generated plan is conditioned on the reference speaker and speaking style attributes. The code generation process can be expressed as the following conditional distribution:
\begin{align}
    \label{eq:q-decoder}
    p(\mathbf{q}_{1:6} \mid \mathcal{P}; \mathbf{p}; \psi) &= p(\mathbf{q}_{1} \mid \mathcal{P}; \mathbf{p}_{1}; \mathcal{F}_{\psi}^1)  \nonumber\\
    &\times \prod_{j=2}^{6} p(\mathbf{q}_{j} \mid \mathbf{q}_{j-1}; \mathbf{p}_{j}; \mathcal{F}_{\psi}^j),
\end{align}
where $\mathbf{q}_j$ and $\mathbf{p}_j$ are the generated target code sequence and prompt code sequence at level $j$, respectively, and $\mathcal{F}_{\psi}^j$ denotes the corresponding decoder FFT block at level $j$. The prior loss $\mathcal{L}_{prior}$ minimizes the negative logarithm of Eq.~\ref{eq:q-decoder}. Although this stage gives the model a stable discrete plan, it is not used as the final acoustic representation.

\begin{figure}[htbp]
    \centering
    \includegraphics[width=0.5\textwidth]{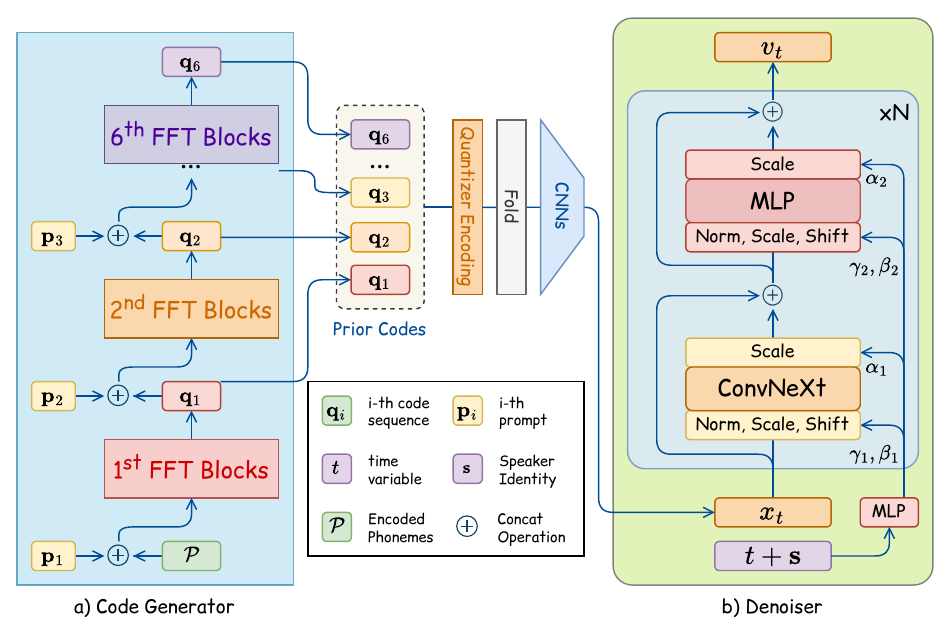}
    \caption{From discrete plan to shortcut latent generation. The Code Generator predicts hierarchical codec tokens conditioned on encoded phonemes and prompt codes. These tokens are embedded and folded into a continuous prior, then refined by the flow-matching Denoiser into a holistic speech latent representation.}
    \label{fig:qdecoder-denoiser}
\end{figure}

\subsection{Shortcut Latent Flow Matching}
\label{sec:shortcut-flow}
The key distinction between the cascade and shortcut approaches lies in how the generated discrete tokens are used. 
In the cascade approach, tokens are converted to speech latents by combining the codewords retrieved from codec codebooks. Such a process is prone to losing fine acoustic details or error accumulation from token prediction errors. In the shortcut approach, token embeddings define the initial point of a continuous generation process. This lets the model retain the alignment and stability of discrete tokens while learning to recover details not captured by the finite codebooks.

Let $\mathbf{x}_{\mathrm{pr}}$ be the continuous prior obtained by embedding and folding $\mathbf{Q}^{1:6}$. We initialize the flow-matching path with a noisy version of this prior:
\begin{equation}
\label{eq:shortcut-prior}
\mathbf{x}'_0 = \mathbf{x}_{\mathrm{pr}} + \tau \epsilon, \qquad \epsilon \sim \mathcal{N}(0,\mathbf{I}),
\end{equation}
where $\tau$ controls the amount of stochasticity around the prior. The interpolation between the prior and the target latent is:
\begin{equation}
\label{eq:shortcut-path}
\mathbf{x}_t = t\mathbf{x}_1 + (1-t)\mathbf{x}'_0, \qquad t \sim \mathcal{U}(0,1).
\end{equation}
The Denoiser predicts the velocity from $\mathbf{x}_t$ to $\mathbf{x}_1$:
\begin{equation}
\label{eq:cfm-loss}
\mathcal{L}_{CFM}(\theta) = \mathbb{E}_{\mathbf{x}_{\mathrm{pr}},\mathbf{x}_1,t}
\left\lVert
\mathbf{v}_{\theta}(\mathbf{x}_t,s,t) - (\mathbf{x}_1 - \mathbf{x}'_0)
\right\rVert^2.
\end{equation}
Since the prior already encodes text alignment and coarse acoustic structure, the Denoiser only needs to perform latent refinement. Accordingly, we replace the multi-head self-attention layers in the DiT-style vector-field estimator \cite{dit} with ConvNeXt blocks, reducing the computation complexity from $\mathcal{O}(L^2 d)$ to $\mathcal{O}(Lk^2 d)$, where $L$ is the sequence length, $k$ is the convolution kernel size, and $d$ is the hidden dimension. The detailed architecture is depicted in Fig.~\ref{fig:qdecoder-denoiser}.

\subsection{Training Objective}
\label{sec:total-loss}
The overall training objective combines discrete planning, rhythm generation, and continuous latent refinement:
\begin{equation}
\mathcal{L}_{total} = \mathcal{L}_{prior} + \mathcal{L}_{dur} + \mathcal{L}_{sil} + \mathcal{L}_{CFM} + \mathcal{L}_{anchor}.
\end{equation}
The anchor loss stabilizes latent refinement by reconstructing the target latent from the predicted velocity:
\begin{equation}
\begin{aligned}
\mathcal{L}_{anchor}
&= \left\lVert \tilde{\mathbf{x}}_1 - \mathbf{x}_1 \right\rVert^2,
\end{aligned}
\end{equation}
\begin{equation}
\begin{aligned}
\tilde{\mathbf{x}}_1
&= \mathbf{x}_t + (1-t)\mathbf{v}_\theta(\mathbf{x}_t, s, t).
\end{aligned}
\end{equation}
The Code Generator and rhythm modules learn the discrete and temporal plan, while the Denoiser learns the shortcut from that plan to fine-grained continuous speech latents.

\section{Experiments}
\begin{table*}[t]
\caption{Main zero-shot TTS results on \textit{LibriSpeech test-clean} using a 3-second prompt. Among synthesized systems, \textbf{bold} and \underline{underlined} values denote the best and second-best results, respectively; ties receive the same formatting. $[\dagger]$: official release; $[\diamond]$: reproduced in our evaluation pipeline; $[\ddagger]$: unofficial checkpoint. Because our method and baselines can differ vastly in data scale and architecture, results should be read with the training-data column. E, VB, LT, GS, and P denote Emilia, VoxBox, LibriTTS, GigaSpeech, and private data, respectively. Results for 1- and 5-second prompts are reported in Appendix Table~\ref{tab:prompt-extremes}.}
\label{tab:overall-results}
\vspace{-0.35cm}
\centering
\resizebox{\textwidth}{!}{%
\begin{tabular}{lccccccccc}
\toprule
\multirow{2}{*}{\textbf{Model}}
& \textbf{Training data}
& \multirow{2}{*}{\textbf{UTMOS} ($\uparrow$)}
& \multirow{2}{*}{\textbf{WER} ($\downarrow$)}
& \multirow{2}{*}{\textbf{SIM-O} ($\uparrow$)}
& \multirow{2}{*}{\textbf{SIM-R} ($\uparrow$)}
& \multicolumn{2}{c}{\textbf{F0}}
& \multicolumn{2}{c}{\textbf{Energy}} \\
\cmidrule(lr){7-8}\cmidrule(lr){9-10}
& (hours) & & & & & \textbf{Accuracy} ($\uparrow$) & \textbf{RMSE} ($\downarrow$) & \textbf{Accuracy} ($\uparrow$) & \textbf{RMSE} ($\downarrow$) \\
\midrule
Ground-truth & -- & 4.09 & 0.02 & -- & -- & -- & -- & -- & -- \\
\midrule
VoxCPM (0.5B) $[\dagger]$
& P (1.8M) & \underline{4.12} & 0.09 & \textbf{0.70} & \textbf{0.70} & 0.78 & 15.92 & 0.68 & 0.009 \\
MaskGCT $[\dagger]$
& E (100k) & 3.79 & 0.10 & \underline{0.67} & \underline{0.69} & 0.78 & 14.90 & \textbf{0.75} & \underline{0.007} \\
Spark-TTS $[\dagger]$
& VB (100k) & \textbf{4.31} & 0.10 & 0.57 & \underline{0.69} & \underline{0.87} & \underline{10.02} & 0.52 & 0.014 \\
VoiceCraft $[\dagger]$
& GS (9k) & 3.55 & 0.18 & 0.51 & 0.45 & 0.78 & 17.22 & 0.44 & 0.010 \\
NaturalSpeech 2 $[\ddagger]$
& LT (585) & 2.38 & 0.09 & 0.31 & 0.38 & 0.80 & 15.62 & 0.25 & 0.020 \\
VALL-E $[\diamond]$
& LT (500) & 3.68 & 0.19 & 0.40 & 0.48 & 0.75 & 21.66 & 0.36 & 0.020 \\
F5-TTS $[\dagger]$
& E (100k) & 3.87 & \textbf{0.04} & 0.64 & -- & 0.82 & 12.64 & 0.61 & 0.014 \\
ZipVoice $[\diamond]$
& LT (500) & 4.01 & 0.14 & 0.59 & -- & 0.79 & 16.35 & 0.44 & 0.014 \\
OZSpeech $[\dagger]$
& LT (500) & 3.15 & \underline{0.05} & 0.40 & 0.47 & 0.81 & 11.96 & 0.67 & 0.010 \\
\rowcolor{verylightgray}
\textbf{Rhythm-TTS (Ours)}
& LT (500) & 3.79 & \textbf{0.04} & 0.51 & 0.59 & \textbf{0.92} & \textbf{6.90} & \underline{0.72} & \textbf{0.006} \\
\bottomrule
\end{tabular}%
}
\end{table*}


\subsection{Experimental Setup}
\label{sec:experimental-setup}
\noindent\textbf{Data and evaluation conditions}. Rhythm-TTS is trained on 500 hours of \textit{LibriTTS} \cite{libritts} and evaluated zero-shot on unseen speakers from \textit{LibriSpeech test-clean} \cite{librispeech}. We use 3-second prompts in the main paper; 1- and 5-second results appear in Appendix~\ref{appendix:prompt-length}.

\noindent\textbf{Protocol and metrics}. Fidelity metrics use one synthesis run per input text-prompt pair, while rhythm diversity is measured as the mean pairwise distance across three independent syntheses. We report WER, UTMOS, SIM-O/SIM-R, pitch and energy transfer, model size, NFE, and RTF. Rhythm diversity is measured by DS-WED \cite{dswed} and the proposed diversity-to-WER ratio, defined as $\text{DWR} = \text{DS-WED}/\text{WER}$. Since DS-WED may be artificially inflated by transcription errors (i.e., higher WER), DWR normalizes diversity by recognition error to better isolate rhythm variation. Additional details are provided in Appendix~\ref{appendix:metrics}.


\subsection{Main Results}
\noindent\textbf{Overall results}. Table~\ref{tab:overall-results} shows the overall results at 3-second prompt length. Rhythm-TTS achieves the lowest WER (0.04) and the smallest pitch and energy reconstruction errors, with $\mathrm{F0}_{\mathrm{RMSE}}=6.90$ and $\mathrm{Energy}_{\mathrm{RMSE}}=0.006$. Among systems trained on the same training data scale (500-585 hours of LibriTTS), it also attains strong naturalness and speaker similarity, reaching a UTMOS of 3.79 and the highest available SIM-R of 0.59. Compared with all baselines, Rhythm-TTS achieves the most faithful prompt-conditioned realization of pitch and energy, obtaining the best $\mathrm{F0}_{\mathrm{ACC}}$, $\mathrm{F0}_{\mathrm{RMSE}}$, and $\mathrm{Energy}_{\mathrm{RMSE}}$, while ranking second in $\mathrm{Energy}_{\mathrm{ACC}}$. Notably, it reduces $\mathrm{F0}_{\mathrm{RMSE}}$ by 31\% relative to the next best baseline.
Although large-scale systems such as Spark-TTS, VoxCPM, and MaskGCT remain stronger on selected metrics, they are trained on orders of magnitude more training data (100k--1.8M hours of speech) and therefore serve primarily as high-capacity reference points. Overall, the matched-data results support the proposed design: discrete tokens preserve linguistic stability and alignment, while shortcut continuous generation improves the acoustic realization of speaker identity, prosody, and naturalness.

\begin{table}[t]
\caption{Core controlled comparison between cascade discrete-token reconstruction and the proposed shortcut continuous-latent generation across prompt lengths. Both variants use the same predicted codec-token plan, so the comparison isolates how that plan is converted into the final acoustic latent. For each prompt length, \textbf{bold} and \underline{underlined} values denote the best and second-best results, respectively}
\label{tab:shortcut-ablation}
\vspace{-0.35cm}
\centering
\resizebox{\columnwidth}{!}{%
\begin{tabular}{lcccc}
\toprule
\textbf{Approach} & \textbf{UTMOS} ($\uparrow$) & \textbf{WER} ($\downarrow$) & \textbf{SIM-O} ($\uparrow$) & \textbf{SIM-R} ($\uparrow$) \\
\midrule
\multicolumn{5}{c}{\textit{1-second prompt}} \\
\midrule
Cascade & \underline{3.13} & \textbf{0.04} & \underline{0.28} & \underline{0.33} \\
\rowcolor{verylightgray}
Shortcut & \textbf{3.52} & \underline{0.05} & \textbf{0.37} & \textbf{0.42} \\
\midrule
\multicolumn{5}{c}{\textit{3-second prompt}} \\
\midrule
Cascade & \underline{3.17} & \textbf{0.04} & \underline{0.36} & \underline{0.44} \\
\rowcolor{verylightgray}
Shortcut & \textbf{3.79} & \textbf{0.04} & \textbf{0.51} & \textbf{0.59} \\
\midrule
\multicolumn{5}{c}{\textit{5-second prompt}} \\
\midrule
Cascade & \underline{3.18} & \textbf{0.04} & \underline{0.39} & \underline{0.48} \\
\rowcolor{verylightgray}
Shortcut & \textbf{3.87} & \textbf{0.04} & \textbf{0.51} & \textbf{0.59} \\
\bottomrule
\end{tabular}%
}
\end{table}

\noindent\textbf{Shortcut generation}. Table~\ref{tab:shortcut-ablation} 
isolates the effect of the reconstruction method: codebook lookup and aggregation (cascade approach) and our proposed shortcut continuous-latent generation approach. Replacing codebook reconstruction with shortcut generation significantly improves all the auditory metrics (UTMOS, SIM-O, SIM-R) while maintaining a nearly unchanged, remarkably low WER across 1-, 3-, and 5-second prompt lengths. For example, at 3-second prompt length, the UTMOS score increases by 0.39 (12\%), 0.62 (20\%), and 0.69 (22\%), respectively.  Because the lexical plan and alignment are identical, the improvement is attributable to the acoustic realization rather than token prediction. Moreover, the gap widens with longer prompts: cascade UTMOS changes only from 3.13 to 3.18 between 1 and 5 seconds, while shortcut generation improves from 3.52 to 3.87. These results support the central claim that codec tokens provide a stable content and alignment prior, but conventional codebook reconstruction forms a granularity bottleneck; our shortcut continuous latent generator better exploits prompt context to recover speaker-dependent spectral details, naturalness, and fine-grained identity cues.

\begin{table}[!t]
\caption{Comparison of run-to-run rhythm diversity across models using a 3-second prompt. Synthesis is repeated three times for each text-prompt pair, and DS-WED is computed over all pairwise synthesis combinations. DS-WED and WER are reported as mean $\pm$ standard deviation over evaluation items; DWR is computed from the corresponding mean DS-WED and mean WER. Bold and underlined values denote the best and second-best mean results, respectively. Provenance symbols follow
Table~\ref{tab:overall-results}.}
\label{tab:temp-results}
\centering
\resizebox{\columnwidth}{!}{%
\begin{tabular}{clccc}
\toprule
\textbf{Approach}
& \textbf{Model}
& \textbf{DS-WED} ($\uparrow$)
& \textbf{WER} ($\downarrow$)
& \textbf{DWR} ($\uparrow$) \\
\midrule
\multirow{2}{*}{AR}
& VoxCPM $[\dagger]$ & 102.78±46.446 & 0.09±0.135 & 1142 \\
& Spark-TTS $[\dagger]$ & 0.0007±0.027 & 0.10±0.136 & 0.0077 \\
& VoiceCraft $[\dagger]$ & 108.45±94.955 & 0.18±0.164 & 603 \\
& VALL-E $[\diamond]$ & \textbf{227.51}±179.135 & 0.19±0.2300 & 1197 \\
\midrule
\multirow{9}{*}{NAR}
& FastSpeech2 $[\ddagger]$ & 0.0±0.0 & \textbf{0.04}±0.071 & 0 \\
& Matcha-TTS $[\dagger]$ & 34.84±28.705& \textbf{0.04}±0.060 & 871 \\
& MaskGCT $[\dagger]$ & \underline{178.44}±125.558 & 0.09±0.157 & 1982 \\
& NaturalSpeech 2 $[\ddagger]$ & 77.75±24.403 & 0.09±0.102 & 864 \\
& F5-TTS $[\dagger]$ & 92.94±68.0 & \textbf{0.04}±0.080 & \underline{2324} \\
& ZipVoice $[\diamond]$ & 164.51±108.88 & 0.14±0.142 & 1175 \\
& OZSpeech $[\dagger]$ & 18.88±7.168& \underline{0.05}±0.078 & 338 \\
\cmidrule(lr){2-5}
& $\text{Rhythm-TTS}^{\text{Fixed}}$ & 41.76±33.271& \textbf{0.04}±0.080 & 1044 \\
\rowcolor{verylightgray}
& Rhythm-TTS & 112.22±34.875 & \textbf{0.04}±0.076 & \textbf{2806} \\
\bottomrule
\end{tabular}%
}
\end{table}

\begin{figure*}[t]
    \centering
    \includegraphics[width=1.0\textwidth]{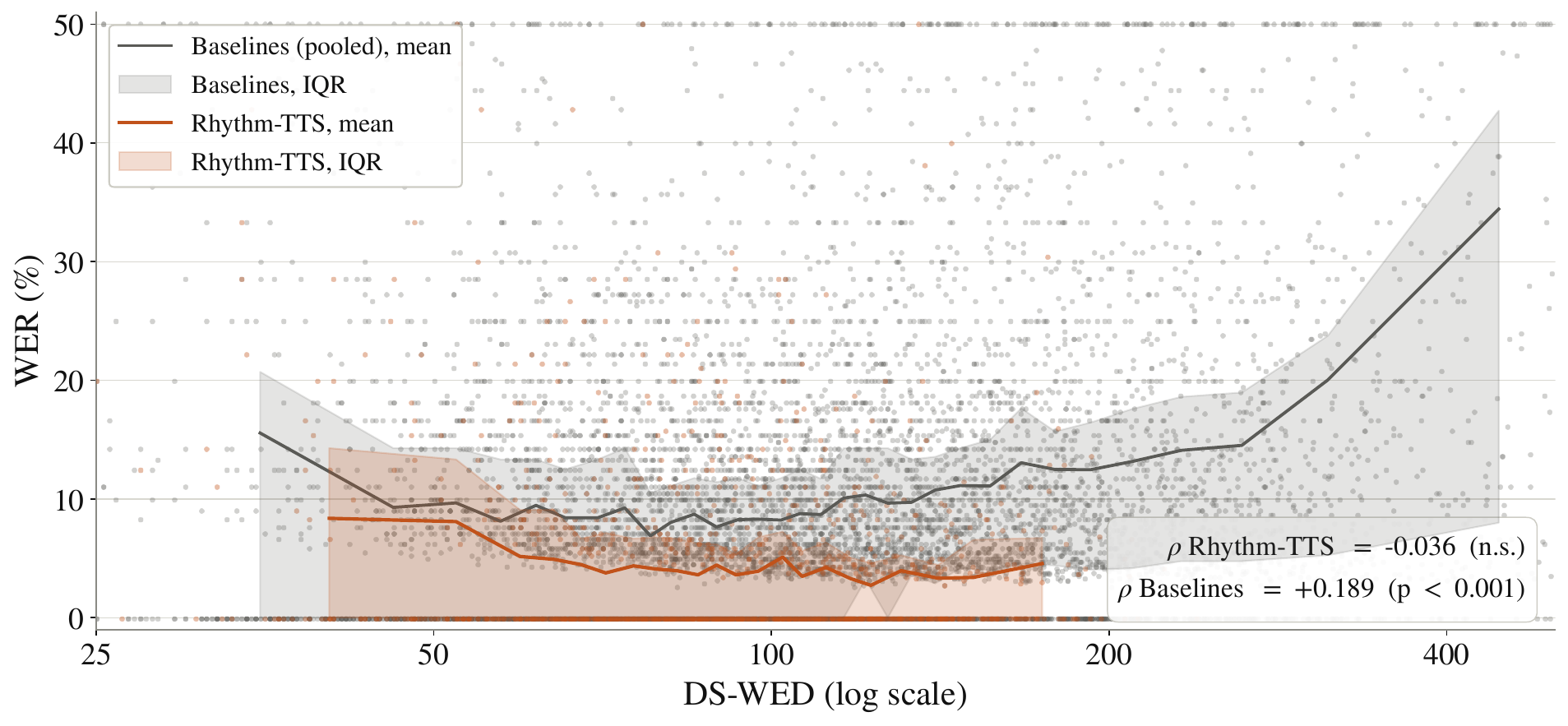}
    \caption{Relationship between run-to-run temporal diversity (DS-WED) and recognition error (WER, \%). Samples are grouped into equal-population quantile bins separately for Rhythm-TTS (22 bins) and the pooled baselines (32 bins). Lines denote mean WER, shaded regions denote the interquartile range, and faint points represent individual synthesis cases. The DS-WED axis is shown on a logarithmic scale. Spearman correlation is non-significant for Rhythm-TTS ($\rho=-0.036$, $p=0.20$), whereas the pooled baselines exhibit a significant positive association ($\rho=0.189$, $p<0.001$). IQR denotes interquartile range.}
    \label{fig:dswed_vs_wer}
\end{figure*}

\noindent\textbf{Run-to-run rhythm diversity}. Table~\ref{tab:temp-results} shows the rhythm diversity across repeated synthesis runs. While DS-WED is an appealing metric to measure this diversity, it can  be inflated by large WER. To decouple such inflation, we propose a new metric DWR, which is simply DS-WED normalized by WER. VALL-E and MaskGCT obtain the largest mean DS-WED values (227.51 and 178.44), but also have higher mean WERs (0.19 and 0.09) and substantially larger DS-WED dispersion. Rhythm-TTS reaches $112.22\pm34.88$ DS-WED at a mean WER of 0.04, producing the best DWR (2806). It increases mean DS-WED by $2.69\times$ over
$\text{Rhythm-TTS}^{\text{Fixed}}$, which is $\text{Rhythm-TTS}$ but with durations kept unchanged across runs, without changing mean WER, and outperforms F5-TTS in both DS-WED and DWR at the same mean WER. The result also clarifies that probabilistic acoustic flow matching alone is insufficient: Matcha-TTS and the $\text{Rhythm-TTS}^{\text{Fixed}}$ remain at 34.84 and 41.76 DS-WED, respectively. Explicit duration and silence sampling is therefore responsible for moving stochasticity into the variables that directly determine pacing and pause structure. Note that here F5-TTS is the official checkpoint trained on 100k hours of speech, which is different from the checkpoint in Table 1 which is reproduced on the same data scale with ours for fair comparison. 
 
\noindent\textbf{Sample-level diversity--intelligibility behavior}.
Fig.~\ref{fig:dswed_vs_wer} complements the aggregated results by examining individual synthesis cases. Across the observed Rhythm-TTS support, the binned mean-WER curve remains approximately flat and the sample-level correlation between DS-WED and WER is non-significant ($\rho=-0.036$, $p=0.20$). In contrast, the pooled-baseline WER curve rises toward the high-diversity region and its interquartile range widens, yielding a noticeable positive correlation ($\rho=0.189$, $p<0.001$). This pattern suggests that Rhythm-TTS changes temporal realization without systematically increasing lexical error, whereas extreme diversity among the pooled baselines is more often accompanied by recognition errors. The comparison remains descriptive: Rhythm-TTS has limited support above DS-WED$\approx240$, and the pooled set combines models with different architectures and training scales.

\subsection{Ablation Study}
\begin{table*}[!th]
\caption{Comparison of model size and latency for 3-second audio prompt length.
The \textbf{\#Params} column reports the total parameters required for
end-to-end synthesis, where the first number denotes the trainable zero-shot
generation model and the second denotes the frozen neural codec or vocoder. \textbf{Bold} and \underline{underlined} values denote the best and second-best comparable results, respectively. Provenance symbols follow Table~\ref{tab:overall-results}.}
\label{tab:size-rtf-results}
\centering
\resizebox{\textwidth}{!}{%
\begin{tabular}{llccccc}
\toprule
\textbf{Model}
& \textbf{\#Params}
& \textbf{NFE} ($\downarrow$)
& \textbf{RTF} ($\downarrow$)
& \textbf{UTMOS} ($\uparrow$)
& \textbf{WER} ($\downarrow$)
& \textbf{SIM-O} ($\uparrow$) \\
\midrule
VoxCPM (0.5B) $[\dagger]$ & 652M + 75M & \underline{10} & 0.67 & \underline{4.12} & 0.09 & \textbf{0.70} \\
MaskGCT $[\dagger]$\textsuperscript{*} & 1.4B + 215M & 25 + 45 & 1.27 & -- & -- & -- \\
Spark-TTS $[\dagger]$ & 507M + 156M BiCodec & -- & 1.06 & \textbf{4.31} & 0.10 & 0.57 \\
VoiceCraft $[\dagger]$ & 830M + 14M EnCodec & -- & 1.70 & 3.55 & 0.18 & 0.51 \\
NaturalSpeech 2 $[\ddagger]$ & 378M + 14M EnCodec & 200 & 1.66 & 2.38 & 0.09 & 0.31 \\
VALL-E $[\diamond]$ & 594M + 104M SpeechTokenizer & -- & 0.86 & 3.68 & 0.19 & 0.40 \\
F5-TTS $[\diamond]$ & 336M + 13.5M Vocos & 32 & 0.26 & 3.76 & 0.24 & 0.52 \\
ZipVoice $[\diamond]$ & 123M + 13.5M Vocos & 32 & 0.13 & 4.01 & 0.14 & \underline{0.59} \\
OZSpeech $[\dagger]$ & 145M + 102M FACodec & \textbf{1} & \textbf{0.013} & 3.15 & \underline{0.05} & 0.40 \\
\midrule
\multirow{4}{*}{Rhythm-TTS} & \multirow{4}{*}{143M + 102M FACodec} & 16 & \underline{0.016} & 3.72 & \textbf{0.04} & 0.51 \\
& & 32 & 0.028 & 3.77 & \textbf{0.04} & 0.51 \\
& & 64 & 0.040 & 3.79 & \textbf{0.04} & 0.51 \\
& & 128 & 0.073 & 3.80 & \textbf{0.04} & 0.51 \\
\bottomrule
\end{tabular}%
}
\vspace{1mm}
\begin{minipage}{0.99\textwidth}
\footnotesize
$^{*}$MaskGCT uses its own neural codec and employs two sequential
probabilistic generative stages. Consequently, \(25+45\) reports the NFEs
used by the first and second stages, respectively, rather than the sampling
budget of a single generative stage.
\end{minipage}
\end{table*}

\noindent\textbf{Latency--quality trade-off}. Table~\ref{tab:size-rtf-results} compares model size and latency across models. With NFE=16, Rhythm-TTS reaches RTF=0.016, the second-fastest result after one-step OZSpeech (0.013), while achieving WER=0.04 and SIM-O=0.51. It is $8 - 106 \times$ faster than all other baselines under indentical hardware setup. Its 245M end-to-end parameters are also the second-smallest total after ZipVoice and are comparable to OZSpeech. Increasing NFE from 16 to 128 raises UTMOS from 3.72 to 3.80, with unchanged WER and SIM-O, indicating that later solver steps mainly polish perceptual details. NFE=16-32 therefore offers the most favorable operating region when latency matters.

\begin{table}
\caption{Ablation of the probabilistic rhythm modules using 5-second prompts. Here, \bc[1.2]{1} removes sampled silence insertion, and \bc[1.2]{2} replaces sampled phoneme durations with their predicted mean values. \textbf{Bold} and \underline{underlined} values denote the best and second-best results, respectively; ties receive the same formatting.}
\label{tab:rhythm-ablation}
\vspace{-0.4cm}
\begin{center}
\resizebox{\columnwidth}{!}{%
\begin{tabular}{ccccc}
\toprule
\textbf{Approach} & \textbf{UTMOS} ($\uparrow$) & \textbf{WER} ($\downarrow$) &\textbf{SIM-O} ($\uparrow$) & \textbf{SIM-R} ($\uparrow$) \\
\midrule
\bc[1.5]{1} & \underline{3.79} & \underline{0.05} & \underline{0.50} & \underline{0.58} \\
\bc[1.5]{2} & 2.98 & \textbf{0.04} & 0.36 & 0.45 \\
\bc[1.5]{1} \bc[1.5]{2} & 3.11 & 0.06 & 0.36 & 0.45 \\
\midrule
\rowcolor{verylightgray}
Ours & \textbf{3.87} & \textbf{0.04} & \textbf{0.51} & \textbf{0.59} \\
\bottomrule
\end{tabular}
}
\end{center}
\vspace{-.5cm}
\end{table}

\noindent\textbf{Ablation of probabilistic rhythm modules}. Table~\ref{tab:rhythm-ablation} shows that sampled duration is the dominant rhythm component. Replacing durations with their predicted means reduces UTMOS from 3.87 to 2.98 and SIM-O/SIM-R from 0.51/0.59 to 0.36/0.45, despite preserving a WER of 0.04. Removing sampled silence is less harmful but still degrades UTMOS to 3.79 and increases WER to 0.05. Disabling both modules yields UTMOS=3.11 and WER=0.06, demonstrating a non-additive interaction between pause placement and neighboring phoneme durations. Thus, the two modules should be interpreted jointly: sampled duration provides the primary timing variation, while sampled silence complements it to produce coherent pacing. Notably, our probabilistic rhythm modules account for less than 1.4\% of the total model parameters, making them a highly favorable trade-off. Only the full model jointly achieves the best naturalness, intelligibility, and speaker similarity. 

\section{Conclusion}
We presented Rhythm-TTS, a zero-shot TTS framework that combines stable discrete speech planning with continuous acoustic generation and explicit rhythm modeling. Instead of using codec tokens as the final speech representation, Rhythm-TTS treats them as an intermediate semantic-acoustic prior and reconstructs continuous speech latents through shortcut flow matching. This design preserves the linguistic stability and alignment benefits of token-based modeling while improving naturalness and speaker similarity. In addition, the Probabilistic Rhythm Planner introduce meaningful variation in speaking rate, phoneme timing, and pause structure without compromising intelligibility.
Our experiments show that shortcut continuous latent generation consistently outperforms conventional codebook lookup and aggregation across different prompt lengths, while explicit temporal sampling provides a better diversity-intelligibility trade-off than relying on stochastic acoustic generation alone. 
These results suggest that separating speech synthesis into discrete planning, temporal realization, and continuous acoustic refinement is an effective direction for building stable and efficient zero-shot TTS systems. 

\section*{Limitations}
Rhythm-TTS is currently evaluated mainly on LibriTTS and LibriSpeech test-clean, so its generalization to multilingual, noisy, emotional, conversational, or singing speech remains to be established. The framework still depends on an external neural codec and therefore may inherit codec-specific artifacts or representation limits. Although shortcut latent modeling bypasses the codebook reconstruction path during synthesis, the codec still defines the latent space and final waveform decoder. Finally, rhythm diversity is evaluated with automatic timing metrics; future work should include controlled human studies to better connect duration and silence diversity with perceived naturalness.

\section*{Potential Risks}
\begin{itemize}
    \item \textbf{Voice cloning misuse:} Because \textit{Rhythm-TTS} is a zero-shot TTS system that can mimic speaker identity, prosody, and other traits from short speech prompts, it could be misused to imitate a real person’s voice without consent. 
    \item \textbf{Deceptive “human-like” audio.} Zero-shot TTS can produce speech that is \textit{nearly indistinguishable} from natural speech, which increases the risk of scams, fraud, and misinformation delivered via synthetic audio. 
    \item \textbf{Legal and Copyright Challenges:} Some voices may be protected by copyright, trademark, or rights of publicity, so using them without authorization could violate these protections and trigger legal disputes.
\end{itemize}

\section*{Acknowledgements}
The majority of this work was conducted at the Quantum AI \& Cyber Security Institute, which served as the primary sponsor of this research. We gratefully acknowledge the Institute for its generous support and for providing the research environment and resources that made this work possible.

\bibliography{custom}

\appendix
\label{sec:appendix}
\section{Appendix Roadmap}
The appendix is organized around common reader questions. Appendix~\ref{appendix:implement} specifies the rhythm-sampling procedure. Appendix~\ref{appendix:prompt-length} reports the 1- and 5-second prompt conditions omitted from Table~\ref{tab:overall-results} and analyzes prompt sensitivity. Appendix~\ref{appendix:detailed-ablation} examines the model-capacity trade-off, complementing the main-text rhythm-module ablation. Subsequent sections study NFE and prior-noise scaling. Finally, Appendix~\ref{appendix:protocol} documents evaluation conventions, metrics, and baseline provenance.

\section{Implementation Details}
\label{appendix:implement}
Algorithm~\ref{alg:dur-sampling} summarizes joint duration and silence sampling and makes explicit how the continuous flow outputs are converted into frame counts.
\begin{algorithm}[t]
\caption{Sampling phoneme durations and optional silences}
\label{alg:dur-sampling}
\begin{algorithmic}[1]
\Require Encoded phonemes $\mathbf{p}_{1:L}$ with $\mathbf{p}_{1}=\texttt{[SIL]}$, sampling steps $N$, and $\Delta t=1/N$
\Ensure Expanded phoneme sequence $\mathbf{p}_{\mathrm{expand}}$
\State Sample $\mathbf{d}^{(0)},\mathbf{s}^{(0)}\sim\mathcal{N}(\mathbf{0},\mathbf{I})$, where $\mathbf{d}^{(0)},\mathbf{s}^{(0)}\in\mathbb{R}^{L}$
\For{$n=0$ \textbf{to} $N-1$}
    \State $t_n \gets n/N$
    \State $\mathbf{d}^{(n+1)} \gets \mathbf{d}^{(n)} + \Delta t\,\mathcal{D}_{\phi}(\mathbf{d}^{(n)},\mathbf{p}_{1:L},t_n)$
    \State $\mathbf{s}^{(n+1)} \gets \mathbf{s}^{(n)} + \Delta t\,\mathcal{S}_{\xi}(\mathbf{s}^{(n)},\mathbf{p}_{1:L},t_n)$
\EndFor
\State $\mathbf{p}_{\mathrm{expand}} \gets \{\}$; $p_{\mathrm{sil}}\gets\mathbf{p}_1$
\For{$i=1$ \textbf{to} $L$}
    \If{$i=1$}
        \State $D_i\gets 0$
    \Else
        \State $D_i\gets\max\!\left(\operatorname{round}(\exp(d_i^{(N)})),1\right)$
    \EndIf
    \State $S_i\gets\max\!\left(\operatorname{round}(\exp(s_i^{(N)})),0\right)$
    \State Append $D_i$ copies of $p_i$ to $\mathbf{p}_{\mathrm{expand}}$
    \State Append $S_i$ copies of $p_{\mathrm{sil}}$ to $\mathbf{p}_{\mathrm{expand}}$
\EndFor
\end{algorithmic}
\end{algorithm}

\section{Prompt-Length Robustness}
\label{appendix:prompt-length}
Table~\ref{tab:prompt-extremes} reports the short- and long-prompt conditions removed from the main comparison. These settings serve different diagnostic purposes: the 1-second condition stresses inference from sparse speaker evidence, whereas the 5-second condition tests whether additional context continues to improve speaker and acoustic estimation.

\begin{table*}[t]
\caption{Extended zero-shot TTS results on \textit{LibriSpeech test-clean} using 1- and 5-second prompts. Within each prompt block, \textbf{bold} and \underline{underlined} values denote the best and second-best synthesized results, respectively; ties receive the same formatting. Provenance symbols follow Table~\ref{tab:overall-results}.}
\label{tab:prompt-extremes}
\centering
\resizebox{\textwidth}{!}{%
\begin{tabular}{lccccccccc}
\toprule
\multirow{2}{*}{\textbf{Model}}
& \textbf{Training data}
& \multirow{2}{*}{\textbf{UTMOS} ($\uparrow$)}
& \multirow{2}{*}{\textbf{WER} ($\downarrow$)}
& \multirow{2}{*}{\textbf{SIM-O} ($\uparrow$)}
& \multirow{2}{*}{\textbf{SIM-R} ($\uparrow$)}
& \multicolumn{2}{c}{\textbf{F0}}
& \multicolumn{2}{c}{\textbf{Energy}} \\
\cmidrule(lr){7-8}\cmidrule(lr){9-10}
& (hours) & & & & & \textbf{Accuracy} ($\uparrow$) & \textbf{RMSE} ($\downarrow$) & \textbf{Accuracy} ($\uparrow$) & \textbf{RMSE} ($\downarrow$) \\
\midrule
\multicolumn{10}{c}{\textbf{1-second prompt}} \\
\midrule
VoxCPM (0.5B) $[\dagger]$ & P (1.8M) & \underline{4.06} & \underline{0.10} & \textbf{0.46} & \underline{0.45} & 0.60 & 31.22 & 0.58 & 0.014 \\
MaskGCT $[\dagger]$ & E (100k) & 2.82 & 0.82 & 0.33 & 0.33 & 0.60 & 40.91 & \underline{0.59} & 0.015 \\
Spark-TTS $[\dagger]$ & VB (100k) & \textbf{4.12} & 0.14 & \underline{0.38} & \textbf{0.46} & \textbf{0.85} & \underline{13.81} & 0.47 & 0.016 \\
VoiceCraft $[\dagger]$ & GS (9k) & 3.45 & 0.16 & 0.31 & 0.24 & 0.61 & 31.57 & 0.52 & \underline{0.010} \\
NaturalSpeech 2 $[\ddagger]$ & LT (585) & 2.12 & 0.12 & 0.20 & 0.21 & \underline{0.69} & 26.48 & 0.39 & 0.020 \\
VALL-E $[\diamond]$ & LT (500) & 3.61 & 0.21 & 0.24 & 0.28 & 0.55 & 37.87 & 0.40 & 0.020 \\
F5-TTS $[\ddagger]$ & E (100k) & 3.78 & 0.11 & 0.38 & -- & 0.68 & 26.12 & 0.45 & 0.029 \\
ZipVoice $[\diamond]$ & LT (500) & 3.48 & 0.36 & 0.32 & -- & 0.55 & 35.81 & 0.41 & 0.021 \\
OZSpeech $[\dagger]$ & LT (500) & 3.17 & \textbf{0.05} & 0.30 & 0.33 & 0.62 & 27.70 & 0.49 & 0.020 \\
\rowcolor{verylightgray}
\textbf{Rhythm-TTS (Ours)} & LT (500) & 3.52 & \textbf{0.05} & 0.37 & 0.42 & \textbf{0.85} & \textbf{13.64} & \textbf{0.70} & \textbf{0.007} \\
\midrule
\multicolumn{10}{c}{\textbf{5-second prompt}} \\
\midrule
VoxCPM (0.5B) $[\dagger]$ & P (1.8M) & \underline{4.16} & 0.08 & \textbf{0.76} & \underline{0.76} & 0.84 & 12.48 & 0.69 & 0.008 \\
MaskGCT $[\dagger]$ & E (100k) & 3.88 & 0.08 & \underline{0.74} & \textbf{0.77} & 0.82 & 11.73 & \textbf{0.76} & \textbf{0.005} \\
Spark-TTS $[\dagger]$ & VB (100k) & \textbf{4.33} & 0.11 & 0.61 & 0.74 & \underline{0.91} & \underline{9.95} & 0.50 & 0.013 \\
VoiceCraft $[\dagger]$ & GS (9k) & 3.58 & 0.19 & 0.56 & 0.51 & 0.81 & 14.48 & 0.46 & 0.010 \\
NaturalSpeech 2 $[\ddagger]$ & LT (585) & 2.33 & 0.09 & 0.35 & 0.44 & 0.84 & 13.13 & 0.28 & 0.020 \\
VALL-E $[\diamond]$ & LT (500) & 3.72 & 0.19 & 0.46 & 0.55 & 0.79 & 18.20 & 0.41 & 0.010 \\
F5-TTS $[\ddagger]$ & E (100k) & 3.91 & \textbf{0.03} & 0.70 & -- & 0.87 & 9.79 & 0.61 & 0.012 \\
ZipVoice $[\diamond]$ & LT (500) & 3.97 & 0.14 & 0.66 & -- & 0.82 & 12.84 & 0.42 & 0.014 \\
OZSpeech $[\dagger]$ & LT (500) & 3.15 & 0.05 & 0.39 & 0.48 & 0.83 & 12.05 & 0.67 & 0.010 \\
\rowcolor{verylightgray}
\textbf{Rhythm-TTS (Ours)} & LT (500) & 3.87 & \underline{0.04} & 0.51 & 0.59 & \textbf{0.92} & \textbf{6.38} & \underline{0.74} & \underline{0.006} \\
\bottomrule
\end{tabular}%
}
\end{table*}

\noindent\textbf{Short-prompt robustness}. With only one second of reference speech, Rhythm-TTS ties OZSpeech for the best WER (0.05), ties Spark-TTS for the best F0 accuracy (0.85), and obtains the best F0 RMSE, energy accuracy, and energy RMSE. Its SIM-O/SIM-R of 0.37/0.42 are below the much larger VoxCPM and Spark-TTS systems but are the strongest among the matched 500-hour models for which both similarities are available. This result indicates that the token-derived plan remains stable even when speaker evidence is sparse.

\noindent\textbf{Context gain and saturation}. Increasing the prompt from 1 to 3 seconds improves Rhythm-TTS UTMOS from 3.52 to 3.79, SIM-O from 0.37 to 0.51, SIM-R from 0.42 to 0.59, and F0 RMSE from 13.64 to 6.90. Extending the prompt to 5 seconds yields only another 0.08 UTMOS, leaves both similarity scores unchanged, and modestly improves acoustic tracking. At 5 seconds, Rhythm-TTS retains the best WER and pitch metrics, whereas Spark-TTS remains best in UTMOS and MaskGCT leads SIM-R and energy metrics. Thus, a 3-second prompt provides most of the identity evidence available on this benchmark; longer prompts mainly refine local acoustic statistics.

\noindent\textbf{Reading cross-scale results}. VoxCPM, MaskGCT, and Spark-TTS use substantially more training data than Rhythm-TTS and retain advantages on some absolute naturalness, similarity, or energy metrics. Their results define high-capacity reference points, whereas the 500-hour LibriTTS rows and controlled ablations are more appropriate for attributing gains to shortcut latent modeling.

\section{Model-Capacity Ablation}
\label{appendix:detailed-ablation}
We further test whether the proposed behavior is preserved under a substantially smaller backbone.

\begin{table}
\caption{Comparison of Rhythm-TTS Base (143M trainable parameters) and Small (76M), both trained on 500 hours of LibriTTS and evaluated on \textit{LibriSpeech test-clean} with NFE=128 and $\tau=0.3$. Within each prompt block, \textbf{bold} and \underline{underlined} values denote the best and second-best results, respectively; ties receive the same formatting.}
\label{tab:model-size-results}
\vspace{-0.4cm}
\begin{center}
\resizebox{\columnwidth}{!}{%
\begin{tabular}{lcccc}
\toprule
\textbf{Model Size} & \textbf{UTMOS} ($\uparrow$) & \textbf{WER} ($\downarrow$) &\textbf{SIM-O} ($\uparrow$) & \textbf{SIM-R} ($\uparrow$) \\
\midrule
\multicolumn{5}{c}{\textit{1-second prompt}} \\
\midrule
\rowcolor{verylightgray}
Base & \textbf{3.53} & \underline{0.05} & \textbf{0.37} & \textbf{0.42} \\
Small & \underline{3.37} & \textbf{0.04} & \textbf{0.37} & \underline{0.41} \\
\midrule
\multicolumn{5}{c}{\textit{3-second prompt}} \\
\midrule
\rowcolor{verylightgray}
Base & \textbf{3.80} & \textbf{0.04} & \textbf{0.48} & \textbf{0.55} \\
Small & \underline{3.58} & \textbf{0.04} & \underline{0.47} & \underline{0.54} \\
\midrule
\multicolumn{5}{c}{\textit{5-second prompt}} \\
\midrule
\rowcolor{verylightgray}
Base & \textbf{3.88} & \textbf{0.04} & \textbf{0.51} & \textbf{0.59} \\
Small & \underline{3.65} & \textbf{0.04} & \underline{0.50} & \underline{0.57} \\
\bottomrule
\end{tabular}
}
\end{center}
\end{table}

\noindent\textbf{Capacity--quality trade-off}. Table~\ref{tab:model-size-results} reduces the trainable backbone from 143M to 76M parameters, a 46.9\% reduction. The Small model preserves WER and loses at most 0.02 in speaker similarity, but UTMOS drops by 0.16, 0.22, and 0.23 for 1-, 3-, and 5-second prompts. At 1 second it even improves WER from 0.05 to 0.04 while matching SIM-O, showing that lexical stability is not the principal capacity bottleneck. The larger model mainly benefits fine acoustic realization, especially as more prompt context becomes available. The Small variant is therefore suitable for resource-constrained deployment, whereas Base is preferable when perceptual naturalness is prioritized.

\section{NFE \& Noise Scaling Analysis}
\begin{table}
\caption{Effect of NFE on \textit{LibriSpeech test-clean} using 5-second prompts with $\tau=0.3$. \textbf{Bold} and \underline{underlined} values denote the best and second-best results, respectively; ties receive the same formatting.}
\label{tab:nfe-effect}
\vspace{-0.4cm}
\begin{center}
\resizebox{\columnwidth}{!}{%
\begin{tabular}{ccccc}
\toprule
\textbf{NFE} & \textbf{UTMOS} ($\uparrow$) & \textbf{WER} ($\downarrow$) &\textbf{SIM-O} ($\uparrow$) & \textbf{SIM-R} ($\uparrow$) \\
\midrule
2 & 3.13 & \textbf{0.03} & 0.47 & 0.54 \\
4 & 3.49 & \underline{0.04} & 0.49 & 0.56 \\
8 & 3.68 & \underline{0.04} & \underline{0.50} & \underline{0.58} \\
16 & 3.79 & \underline{0.04} & \textbf{0.51} & \textbf{0.59}  \\
32 & 3.84 & \underline{0.04} & \textbf{0.51} & \textbf{0.59} \\
64 & 3.87 & \underline{0.04} & \textbf{0.51} & \textbf{0.59} \\
128 & \underline{3.88} & \underline{0.04} & \textbf{0.51} & \textbf{0.59} \\
\rowcolor{verylightgray}
256 & \textbf{3.90} & \underline{0.04} & \textbf{0.51} & \textbf{0.59} \\
\bottomrule
\end{tabular}
}
\end{center}
\end{table}

\noindent\textbf{Number of function evaluations}. Table~\ref{tab:nfe-effect} shows rapid early improvement followed by saturation. Raising NFE from 2 to 16 improves UTMOS from 3.13 to 3.79 and SIM-O/SIM-R from 0.47/0.54 to 0.51/0.59. Although NFE=2 gives the lowest WER (0.03), its markedly poorer naturalness and similarity show that WER alone does not capture acoustic convergence. Beyond 16 evaluations, WER and speaker similarity are unchanged, while UTMOS rises gradually to 3.90 at NFE=256. A $16\times$ increase from 16 to 256 therefore yields only 0.11 UTMOS, placing NFE=16--32 near the knee of the quality--computation curve.

\begin{table}
\caption{Effect of the prior-noise scale $\tau$ on \textit{LibriSpeech test-clean} using 5-second prompts with NFE=64. \textbf{Bold} and \underline{underlined} values denote the best and second-best results, respectively; ties receive the same formatting.}
\label{tab:tau-effect}
\vspace{-0.4cm}
\begin{center}
\resizebox{\columnwidth}{!}{%
\begin{tabular}{ccccc}
\toprule
\textbf{$\tau$} & \textbf{UTMOS} ($\uparrow$) & \textbf{WER} ($\downarrow$) &\textbf{SIM-O} ($\uparrow$) & \textbf{SIM-R} ($\uparrow$) \\
\midrule
0.0 & 3.70 & \textbf{0.04} & \underline{0.50} & \underline{0.58} \\
0.1 & 3.79 & \textbf{0.04} & \textbf{0.51} & \underline{0.58} \\
0.2 & \underline{3.85} & \textbf{0.04} & \textbf{0.51} & \textbf{0.59} \\
\rowcolor{verylightgray}
0.3 & \textbf{3.87} & \textbf{0.04} & \textbf{0.51} & \textbf{0.59}  \\
\hdashline
0.4 & 3.84 & \textbf{0.04} & \underline{0.50} & \textbf{0.59} \\
0.5 & 3.77 & \underline{0.05} & \underline{0.50} & \textbf{0.59} \\
0.6 & 3.67 & \underline{0.05} & 0.49 & \underline{0.58} \\
0.7 & 3.55 & 0.07 & 0.47 & 0.57 \\
0.8 & 3.40 & 0.10 & 0.46 & 0.55 \\
0.9 & 3.21 & 0.12 & 0.43 & 0.53 \\
1.0 & 2.99 & 0.18 & 0.41 & 0.50 \\
\bottomrule
\end{tabular}
}
\end{center}
\end{table}

\noindent\textbf{Noise scaling}. Table~\ref{tab:tau-effect} exhibits a clear optimum around $\tau=0.3$. Relative to no perturbation, $\tau=0.3$ raises UTMOS from 3.70 to 3.87 while retaining the best WER tier and improving similarity. The second-best UTMOS occurs at $\tau=0.2$, showing that a small neighborhood around the codec prior is consistently beneficial. Beyond $\tau=0.4$, quality degrades: at $\tau=1.0$, WER reaches 0.18 and UTMOS falls to 2.99. Moderate noise therefore gives the Denoiser room to recover residual continuous detail, whereas excessive noise destroys the alignment and acoustic structure supplied by the prior. This behavior supports the interpretation of Rhythm-TTS as a local latent refiner rather than an unconditional generator.

\section{Experiment Details}
\label{appendix:protocol}
\subsection{Evaluation Conditions and Reporting Conventions}
Within each prompt-length block, all systems are evaluated on the same text--prompt pairs. Fidelity metrics use one synthesis per pair, whereas DS-WED uses three independent passes of the same pair and averages all pairwise distances. RTF is measured end-to-end on a single NVIDIA A100 80GB GPU and includes the zero-shot generator and frozen waveform reconstruction component. Results from differently scaled training corpora are included for context, but architectural conclusions are based primarily on the matched LibriTTS comparisons and within-model ablations.

The tables report raw values rather than aggregating all criteria into one global score. This is intentional: WER, naturalness, identity similarity, prompt-acoustic matching, latency, and rhythm diversity represent distinct deployment priorities. DWR is therefore treated as a diagnostic companion to DS-WED, not as a replacement for the underlying WER and diversity measurements.

\subsection{Metrics}
\label{appendix:metrics}
\begin{itemize}
    \item \textbf{Real-Time Factor (RTF)}: A key measure of system efficiency, especially for real-time applications, RTF indicates the time needed to generate one second of speech. We evaluate RTF for all models in a complete end-to-end setup on an NVIDIA 80GB A100 GPU.

    \item \textbf{Number of Function Evaluations (NFE)}: This metric counts how many times the model's guiding function (e.g., score or drift function) is calculated during sampling. It’s particularly relevant when the generative process involves solving an ordinary differential equation (ODE), as in score-based generative models using the probability flow ODE method.

    \item \textbf{UTMOS}: A deep learning-based tool for evaluating speech quality by predicting mean opinion scores (MOS). It replaces expensive and time-intensive subjective evaluations, offering predictions that closely match human judgments.

    \item \textbf{SIM-O and SIM-R}: These metrics measure speaker similarity. SIM-O compares synthesized speech to the original prompt, while SIM-R compares it to a reconstructed prompt from FACodec. Both are calculated using cosine similarity of speaker embeddings extracted via WavLM-TDCNN, with values ranging from -1 to 1 (higher values indicate greater similarity).

    \item \textbf{Word Error Rate (WER)}: This evaluates the accuracy of word pronunciation in speech synthesis systems. We use a CTC-based HuBERT model, pre-trained on LibriLight and fine-tuned on LibriSpeech’s 960-hour dataset, to transcribe generated speech and compare it to the text prompt.

    \item \textbf{Acoustic Accuracy \& Error ($\mathbf{F0}_{\mathbf{ACC}}$, $\mathbf{F0}_{\mathbf{RMSE}}$, $\mathbf{EN}_{\mathbf{ACC}}$, $\mathbf{EN}_{\mathbf{RMSE}}$)}: These assess how well synthesized speech aligns with the audio prompt, focusing on pitch ($\mathbf{F0}$) and energy ($\mathbf{EN}$). Accuracy is measured by categorizing F0 and energy into high, normal, and low levels, following methods from PromptTTS and TextrolSpeech. Additionally, Root Mean Square Error (RMSE) quantifies differences in F0 and energy between synthesized speech and prompts.

    \item \textbf{Discretized Speech Weighted Edit Distance (DS-WED)}: This metric measures run-to-run rhythm diversity by comparing duration-and-silence sequences generated from multiple synthesis passes of the same text and prompt. Higher DS-WED indicates stronger variation in phoneme timing and pause structure.

    \item \textbf{Diversity--WER Ratio (DWR)}: We compute $\text{DWR}=\mathrm{DS\mbox{-}WED}/\mathrm{WER}$ to contextualize temporal diversity by intelligibility. A high value indicates substantial timing variation relative to recognition error. Because ratios can become large when WER is small, DWR is always reported together with raw DS-WED and WER and is used as a diagnostic rather than a perceptual quality metric.
\end{itemize}

\subsection{Baselines}
\label{appendix:baselines}
\noindent\textbf{Provenance convention}. We use three compact symbols consistently across all result tables: $[\dagger]$ denotes an official implementation, checkpoint, or released evaluation; $[\diamond]$ denotes a system reproduced in our evaluation pipeline; and $[\ddagger]$ denotes an unofficial or community checkpoint. These symbols identify result provenance rather than comparison fairness. The reported training scale, tokenizer or vocoder, inference configuration, and evaluation protocol should therefore be considered jointly when interpreting cross-system differences.

\noindent\textbf{Coverage}. The following inventory includes every external system appearing in Tables~\ref{tab:overall-results},~\ref{tab:size-rtf-results},~\ref{tab:temp-results}, and~\ref{tab:prompt-extremes}. Controlled conclusions are drawn primarily from systems trained on LibriTTS and from within-model ablations; substantially larger models and datasets are retained as high-capacity reference points.

\begin{itemize}
    \item \textbf{VoxCPM (0.5B) $[\dagger]$}: We use the official 0.5B release. The model is included as a large-scale autoregressive reference and is reported as trained on approximately 1.8M hours of private speech. Because neither its training corpus nor its scale is matched to Rhythm-TTS, its results provide external context rather than a controlled architecture comparison.

    \item \textbf{MaskGCT $[\dagger]$}: We use the official released system trained on the 100k-hour Emilia corpus. MaskGCT is a non-autoregressive masked codec-token baseline and is evaluated in the fidelity, prompt-length, and rhythm-diversity comparisons. Its much larger training set should be considered when comparing absolute naturalness and speaker-similarity scores.

    \item \textbf{Spark-TTS $[\dagger]$} \cite{sparktts}: We use the official implementation\footnote{\url{https://github.com/SparkAudio/Spark-TTS}} and official Spark-TTS-0.5B checkpoint\footnote{\url{https://huggingface.co/SparkAudio/Spark-TTS-0.5B}}, trained on the 100k-hour VoxBox corpus introduced with the model. Spark-TTS uses BiCodec \cite{sparktts} as its speech tokenizer and serves as a large-scale codec-token baseline.

    \item \textbf{VoiceCraft $[\dagger]$} \cite{voicecraft}: We adopt the official implementation and the 830M-parameter TTS-enhanced checkpoint\footnote{\url{https://huggingface.co/pyp1/VoiceCraft/blob/main/830M_TTSEnhanced.pth}}, trained on GigaSpeech \cite{gigaspeech}. VoiceCraft uses EnCodec \cite{encodec}; its autoregressive decoding and large parameter count make it an informative quality--latency reference.

    \item \textbf{NaturalSpeech 2 $[\ddagger]$} \cite{ns2}: We evaluate the publicly available Amphion implementation \cite{amphion} and its LibriTTS checkpoint\footnote{\url{https://huggingface.co/amphion/naturalspeech2_libritts/tree/main/checkpoint}}. Because this is not an official author release, it is marked as an unofficial checkpoint. NaturalSpeech 2 uses EnCodec representations and provides a latent diffusion baseline trained at a scale close to ours.

    \item \textbf{VALL-E $[\diamond]$} \cite{valle}: We reproduce VALL-E with the Amphion framework \cite{amphion}, training it on the same 500-hour LibriTTS corpus used for Rhythm-TTS. It uses SpeechTokenizer \cite{speechtokenizer} and supplies the principal matched-data autoregressive codec-token comparison.

    \item \textbf{F5-TTS $[\diamond]$} \cite{f5tts}: We reproduce the official implementation\footnote{\url{https://github.com/SWivid/F5-TTS}} on the 500-hour LibriTTS corpus following the released training configuration. F5-TTS is a continuous flow-matching baseline using Vocos for waveform synthesis. It is particularly relevant for separating stochastic acoustic generation from explicit duration-and-silence sampling.

    \item \textbf{ZipVoice $[\diamond]$} \cite{zipvoice}: We reproduce ZipVoice on the 500-hour LibriTTS corpus using its official implementation and Vocos backend. ZipVoice combines a Zipformer backbone \cite{zipformer} with flow distillation, providing a compact continuous-generation baseline in the quality, efficiency, prompt-length, and rhythm-diversity comparisons.

    \item \textbf{OZSpeech $[\dagger]$} \cite{oz}: We use the official repository and checkpoint\footnote{\url{https://github.com/ozspeech/OZSpeech}}, trained on the 500-hour LibriTTS corpus. OZSpeech uses FACodec \cite{ns3} and learned-prior-conditioned flow matching with one function evaluation, making it the closest reference for evaluating the efficiency benefit of an informative latent prior.

    \item \textbf{FastSpeech2 $[\ddagger]$}: We use a publicly available unofficial checkpoint solely in the rhythm-diversity comparison. Its deterministic duration prediction and length regulation provide a deliberately low-variance reference: repeated synthesis of the same text--prompt pair should preserve essentially identical temporal structure.

    \item \textbf{Matcha-TTS $[\dagger]$} \cite{matcha-tts}: We use the official released implementation and checkpoint for the rhythm-diversity comparison. Matcha-TTS generates mel-spectrograms with conditional flow matching. It tests whether a probabilistic acoustic-generation objective alone produces temporal diversity when duration and pause structure are not explicitly sampled.
\end{itemize}

\end{document}